\documentstyle[times,epsfig]{aa}

\def\Ms{M_\odot}
\def\rerg{\rm erg}
\def\rs{\rm s}
\def\rs1{\rm s^{-1}}

\def\rcm{\rm cm}
\def\rcm2{\rm cm^{-2}}

\def\flux{\rerg\ \rcm2\ \rs1}

\def\zeta{A_{Fe}}

\def\etal{et al.}
\def\aap{{\it A\&A}}
\def\apj{{\it ApJ }}
\def\apjl{{\it ApJL }}

\def\nat{{\it Nature }}
\def\aas{{\it Astron. Astrophys. Suppl. Ser.}}
\def\mnras{{\it MNRAS}}
\begin{document}

\thesaurus{1
           (13.07.1;
            11.17.4)}
\title{Iron line signatures in X-ray afterglows of GRB by BeppoSAX}

\author{ L. Piro \inst{1} \and
E. Costa \inst{1} \and
M. Feroci \inst{1} \and
G. Stratta \inst{1} \and
F. Frontera \inst{2,3} \and
L. Amati \inst{2} \and
D. Dal Fiume \inst{2} \and
L. A. Antonelli \inst{4,5} \and
J. Heise \inst{6} \and
J. in 't Zand \inst{6} \and
A. Owens \inst{7} \and
A.N. Parmar \inst{7} \and
G. Cusumano \inst{8} \and
M. Vietri \inst{9} \and
G.C. Perola \inst{9}
 }
\offprints{piro@ias.rm.cnr.it}      

\institute{
{Istituto di Astrofisica Spaziale, C.N.R., Roma, Italy} 
\and {Istituto T.E.S.R.E., C.N.R., Bologna, Italy }
\and {Dipartimento di Fisica, Universita' di Ferrara, Italy }
\and {BeppoSAX Science Operation Center, Rome, Italy}
\and {Osservatorio Astronomico di Roma, Italy}
\and {Space Research Organization in the Netherlands, Utrecht, The Netherlands}
\and {Space Science Department of ESA, ESTEC}
\and {Istituto Fisica Cosmica e Appl. Calc. Informatico, C.N.R.,Palermo, Italy}
\and {Dipartimento di Fisica, Universita' Roma Tre, Roma, Italy}
}

\date{Received; Accepted }

\maketitle

\begin{abstract}

We report the possible detection (99.3\% of statistical
significance) of
redshifted Fe iron line emission 
in the X-ray afterglow of 
Gamma-ray burst GRB970508 observed by BeppoSAX. 
Its energy is 
consistent with the redshift of the putative host
galaxy  determined from optical spectroscopy.
In contrast to the fairly clean environment
expected in the merging of two neutron stars,
the observed line properties would imply that
the site of the burst is  embedded in a large
mass of material ($>0.5 \Ms$), consistent with pre-explosion ejecta
of a very massive star.
This material could  be related with the outburst observed
in the afterglow 1 day after the GRB and with 
the spectral variations measured
during this phase. We did not find evidence of Fe line in two 
other GRB with known redshift (GB971214 and GB980613), 
but we note that the upper limits
are of the same order of the intensity measured in GB97508 and that
none of these afterglows shows rebursting activity. 
\end{abstract}

\section{Introduction}

Distance - scale determination of Gamma-ray bursts (GRB) has been 
one of the most important achievements of 
astrophysics in recent years. Accurate and fast localization of the 
prompt and afterglow emission  (\cite{piro98a,Costa97}) by
 BeppoSAX (\cite{P95,B97}) led to the 
identification of optical counterparts (\cite{vpd97})
and ultimately to spectral 
measurements of a redshift (\cite{metzger}). 
While the extragalactic origin of GRB has gathered  solid 
evidence in its 
support, the source of the large  energy implied by
their distance
 is still speculative.

The measurement of X-ray Fe lines emitted 
directly by the GRB or its afterglow
could provide  a direct measurement of the distance and
probe into the nature of the central environment (\cite{pl98,
mr98,bdcl98,ghi98}).  Neutron star -- neutron star merging 
should happen in a fairly clean environment, with line intensities
much below the sensitivity of current experiments. In contrast,
M\'esz\'aros and Rees (1998)  have shown that the
circumburst environment created by the stellar wind before the
explosion of the hypernova could yield a line of substantial
intensity. A similarly favourable situation should be expected in
a variation of the hypernova scenario, -- the SupraNova (\cite{vs98}),
where the GRB is shortly preceded by a supernova explosion with the ejection
of an iron--rich  massive shell. 
It is also conceivable that the impact of the  
 relativistic shell that
produced the original GRB  on these ejecta
could provide an additional 
energy input in the afterglow.

Motivated by these expectations, we have started a detailed analysis
of afterglow spectra to look for the presence of features.
The first and most promising candidate is GB970508.
 It is characterized by a large outbursting event 
during its afterglow phase (\cite{Piro98}) and has
the highest signal to noise ratio of the BeppoSAX GRB afterglows.
Here we summarize the results obtained in this burst
(reported more extensively in Piro \etal 1999)
and present the first results of the analysis in other afterglows.

\section{Possible evidence of Fe line feature in GB970508}

We have searched the X--ray spectrum of GB970508's afterglow for
an iron line, located at the system's redhift (z = 0.835, Bloom et al. 1998); 
we found such a
line with limited statistical significance ($99.3\%$) in the early part 
(first $16\; h$) of the afterglow; the line decreases in the later part
of the observations ($\approx 1$ day after the burst) by at least a factor 2, 
enough to make it undetectable with current apparatus.
Simultaneously with the line disappearance, the X--ray flux both
rises and hardens ($\alpha = 0.4\pm 0.6$, while $\alpha = 1.5\pm 0.6$ before 
the reburst), consistent with the appearance of a new shock. Then,
at the
end of the outburst, the spectrum steepens.

\section{Search of Fe line in other GRB's afterglows}
An extensive analysis is under way to 
search for features in all the X-ray 
afterglows observed by BeppoSAX. We report here  
the results derived in
relatively bright afterglow sources 
(flux in the first part of the observation $>3\times 10^{-13} \flux$) 
which have
a spectroscopic redshift determination, thus allowing
 to fix the energy of the
line.  In the two GRB's that meet these conditions (GRB971214 and GRB980613)
we find upper limits that are roughly consistent with the intensity measured
in GB970508 (Tab.1)

\section{Origin of the line and constraints on the emitting region}
It is most likely that the line of GB970508
is produced by fluorescence and recombination of
Fe atoms ionized by the intense flux of the GRB and its afterglow (\cite{PI},
Lazzati \etal 1999) 
In the early phases of the GRB the radiation field  is 
 so high that iron atoms are 
completely stripped of their electrons: the Compton temperature
is very high and then recombination is not very efficient in producing
line photons. When the flux decreases, 
about $10^4 s$ after the burst, fluorescence becomes an effective 
process. We note, in passing, that the intensity of the line is
therefore not  correlated with the luminosity of the burst:
for example, with a luminosity a factor of 10
{\it larger}, the medium would have remained completely ionized upto
about 1 day after the burst, producing therefore a line with a
{\it lower} intensity.
The minimum mass needed to produce the line is (Piro \etal 1999, Lazzati
\etal 1999)
$M_{min} = 0.5 M_\odot \zeta^{-1}$, where $\zeta$ is the iron abundance 
normalized to the solar value.
From the line variability, intensity and width we deduce that 
this medium should be located at a distance of $\approx 3\times 10^{15}$cm
from the  central source, it is moving with subrelativistic speed,
 it should have a large density
($n>5\times 10^9 cm^{-3}$), and 
it should lie sideways respect to the observer,
otherways it would smear out the short timescale structure of the burst with
Thomson scattering (\cite{bdcl98}).

In order to reach such distance, 
this material must have been pre--ejected by the source originating the burst, 
shortly (perhaps a year, for typical SN expansion speeds) before the burst. 
We stress that these observations contain two coincidences: on the one hand,
this is the only burst in which a reburst and a line have been observed
by BeppoSAX; on
the other, the iron line disappears exactly at the moment of the reburst.
We also point out that a line feature, with a similar significance, has been
found by ASCA in another burst, GRB970828, which
also shows an event of rebursting during the X-ray afterglow
(\cite{yoshida98}).
On the contrary, neither GB971214 nor GB98613 show rebursting
(Costa 1999).

\begin{table}
\caption[] {Iron lines in X-ray afterglows of GRB with known z}
\begin{flushleft}
\begin{tabular}{llllll}
\noalign {\hrule}
\noalign {\medskip}
GRB & $T_{start}^1$ & $F_c^2$ & $I_L^3$ & E.W.$^4$ & z  \\
970508 & 6 & 6 & $5\pm2$ & $3\pm1.4$& 0.835 \\
971214 & 7 & 7 & $<6$ & $<0.5$& 3.14 \\
980613 & 9 & 3 & $<2.5$ & $<2$& 1.096 \\
\noalign {\hrule}
\noalign{\medskip}
\noalign{\noindent
Note: Continuum and line intensities in the period $T_{start}$ to $T_{start}$
+ 30 ksec; Errors and upper limits correspond to 90\% confidence
level for 1 parameter of interest; z of GB971214 from
Kulkarni \etal (1998); z of GB980613 from Djorgovski \etal (1999)}
\noalign{\noindent
$^1$ observation start in hrs from the GRB;
$^2$ F(2-10 keV) in $10^{-13} \flux$;
$^3$ Line intensity in $10^{-5}$ ph cm$^{-2}$ s$^{-1}$;
$^4$ observed E.W. in keV}

\end{tabular}
\end{flushleft}
\end{table}

\begin{figure}
\begin{center}
\epsfig{figure=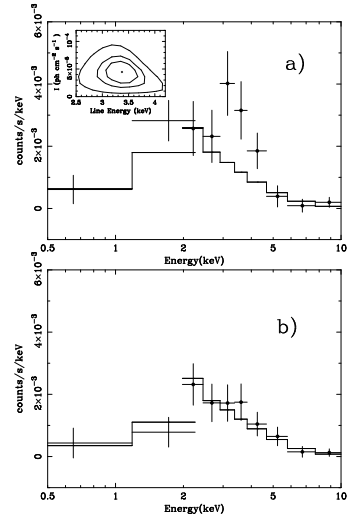,width=6cm}
\caption[]{
The spectra (in detector counts) of the afterglow of GRB970508 
in the first
 (a) and and second part of the
observation (b) fitted with a  power law continuum.
}
\end{center}
\end{figure}

\acknowledgements
We thank  the BeppoSAX team for the support with
observations. BeppoSAX is a program of the Italian space agency
(ASI) with the participation of the Dutch space agency (NIVR).

\end{document}